\documentclass[aps,superscriptaddress,twocolumn]{revtex4}
\usepackage{amsmath}
\usepackage{graphicx}
\usepackage{dcolumn}
\usepackage{verbatim}
\usepackage{times}
\usepackage{subfigure}
\usepackage{bm}
\usepackage{color}
\usepackage[colorlinks,citecolor=blue,linkcolor=blue,dvipdfm]{hyperref}
\usepackage{subfigure}

\usepackage{booktabs}
\usepackage{appendix}

\begin{document}

\title{Switching of topological phase transition from semiconductor to ideal Weyl states in Cu$_2$SnSe$_3$ family of materials}

\author{Huan Li}
\email{lihuan@glut.edu.cn}
\affiliation{College of Physics and Electronic Information Engineering, Guilin University of Technology, Guilin 541004, China}
\affiliation{Key Laboratory of Low-dimensional Structural Physics and Application, Education Department of Guangxi Zhuang Autonomous Region, Guilin 541004, China}

\date{\today}

\begin{abstract}

The exploration of topological phase transition (TPT) mechanisms constitutes a central theme in quantum materials research. Conventionally, transitions between Weyl semimetals (WSMs) and other topological states rely on the breaking of time-reversal symmetry (TRS) or precise manipulation of lattice symmetry, thus constraints the available control strategies and restrict the scope of viable material systems. In this work, we propose a novel mechanism for TPT that operates without TRS breaking or lattice symmetry modification: a class of semiconductors can be directly transformed into an ideal WSM via bandgap closure. This transition is driven by chemical doping, which simultaneously modulates the band gap and enhances spin-orbit coupling (SOC), leading to band inversion between the valence and conduction bands and thereby triggering the TPT. Using first-principles calculations, we demonstrate the feasibility of this mechanism in the Cu$_2$SnSe$_3$ family of materials, where two pairs of Weyl points emerge with energies extremely close to the Fermi level. The bulk Fermi surface becomes nearly point-like, while the surface Fermi surface consists exclusively of Fermi arcs. This symmetry-independent mechanism bypasses the constraints of conventional symmetry-based engineering, and also offers an ideal platform for probing the anomalous transport properties of WSMs.

\end{abstract}

\maketitle

\section{Introduction}

Topological insulators (TIs), Weyl semimetals (WSMs), and Dirac semimetals (DSMs) have emerged as pivotal topics in condensed matter physics and materials science~\cite{Kane05,Wan11,Wang12}. Topological phase transitions (TPTs) among these topological phases are typically marked by abrupt changes in band topology - quantified by topological invariants - and are usually accompanied by energy gap closure at the critical point~\cite{Bansil16}. As a prototypical topological semimetal, the WSM hosts low-energy excitations governed by the Weyl equation, featuring pairs of Weyl points with opposite chirality in the Brillouin zone (BZ). These give rise to distinctive physical phenomena such as surface Fermi arcs and chiral anomalies~\cite{Xu15,Weng15,Taddei22}. The phase transition behavior and its regulatory mechanisms between WSMs and other topological states constitute central scientific challenges in uncovering their topological nature and advancing the rational design of functional materials.

In typical scenarios, TPT from a DSM to a WSM occurs when time-reversal symmetry (TRS) or space inversion symmetry (SIS) is broken, leading to the splitting of a Dirac point into a pair of Weyl points with opposite chirality. For example, magnetic fields, spontaneous magnetization or lowering the crystal symmetry can drive Cd$_3$As$_2$ and CeAsS from DSM to WSM phases, accompanied by pronounced anomalous Hall effects~\cite{Wang13,Li25}. Similarly, transitions between WSMs and insulating phases are commonly achieved via magnetic fields, stacking, strain engineering, or chemical doping, which induce gap opening or closing and alter the topological invariants~\cite{Liu14,Liu16,Armitage17,Chang13,Chen21}. Specific examples include Cr-doped (Bi,Sb)$_2$Te$_3$, in which magnetic ordering splits surface states and generates Weyl Fermi arcs~\cite{Belopolski25}; tension-driven TPT from magnetic WSM to quantum anomalous Hall insulator in Co$_3$Sn$_2$S$_2$~\cite{Liu18,Sun23}; and pressure and field induced TPT from antiferromagnetic insulator to ferromagnetic WSM in Eu$_2$Zn$_2$As$_2$~\cite{Luo23}.
These studies collectively highlight that existing routes to WSMs predominantly depend on symmetry manipulation - particularly TRS breaking or lattice symmetry changes. However, this reliance introduces significant constraints: first, it restricts candidate materials to magnetic or structurally responsive systems, thereby excluding numerous non-magnetic, abundant, and technologically compatible semiconductors; second, symmetry control often demands extreme conditions (e.g., strong magnetic fields, cryogenic temperatures) or intricate synthesis protocols, limiting scalability and practical tunability.

Semiconductors, as foundational electronic materials, generally exhibit topologically trivial band structures. In existing studies, the transition from an insulating state to a WSM typically relies on symmetry modulation through mechanisms such as magnetic ordering, stacking, strain engineering, or disordered doping~\cite{Belopolski25,Sun23,Luo23,Chen15}. While effective, these approaches often entail significant operational complexity. Achieving an efficient and controllable transition from a trivial semiconductor to a WSM without the need to modulate TRS or lattice symmetry has emerged as a research topic of significant scientific importance.
To address this issue, we perform systematic first-principles calculations on the Cu$_2$SnSe$_3$ family of semiconductors, examining their band evolution under spin-orbital coupling (SOC) modulation and chemical doping. We uncover a previously unidentified TPT mechanism: without breaking or modulating time-reversal and lattice symmetry, a trivial semiconductor undergoes a direct transition to a WSM phase through energy gap closure, resulting in the simultaneous emergence of Weyl points at four locations in the BZ. Crucially, these Weyl points lie exceptionally close to the Fermi level, yielding a nearly point-like bulk Fermi surface dominated entirely by Weyl cones, while the surface Fermi surface exhibits only well-defined Fermi arcs - hallmarks of an ideal WSM. This mechanism transcends the constraints of conventional symmetry-driven transitions, opening a new and accessible pathway for topological state engineering, and provides a pristine platform for probing the anomalous transport phenomena associated with Weyl fermions.

\section{Crystal structure and electronic bands}

\begin{figure}[tbp]
\hspace{-0cm} \includegraphics[totalheight=1.5in]{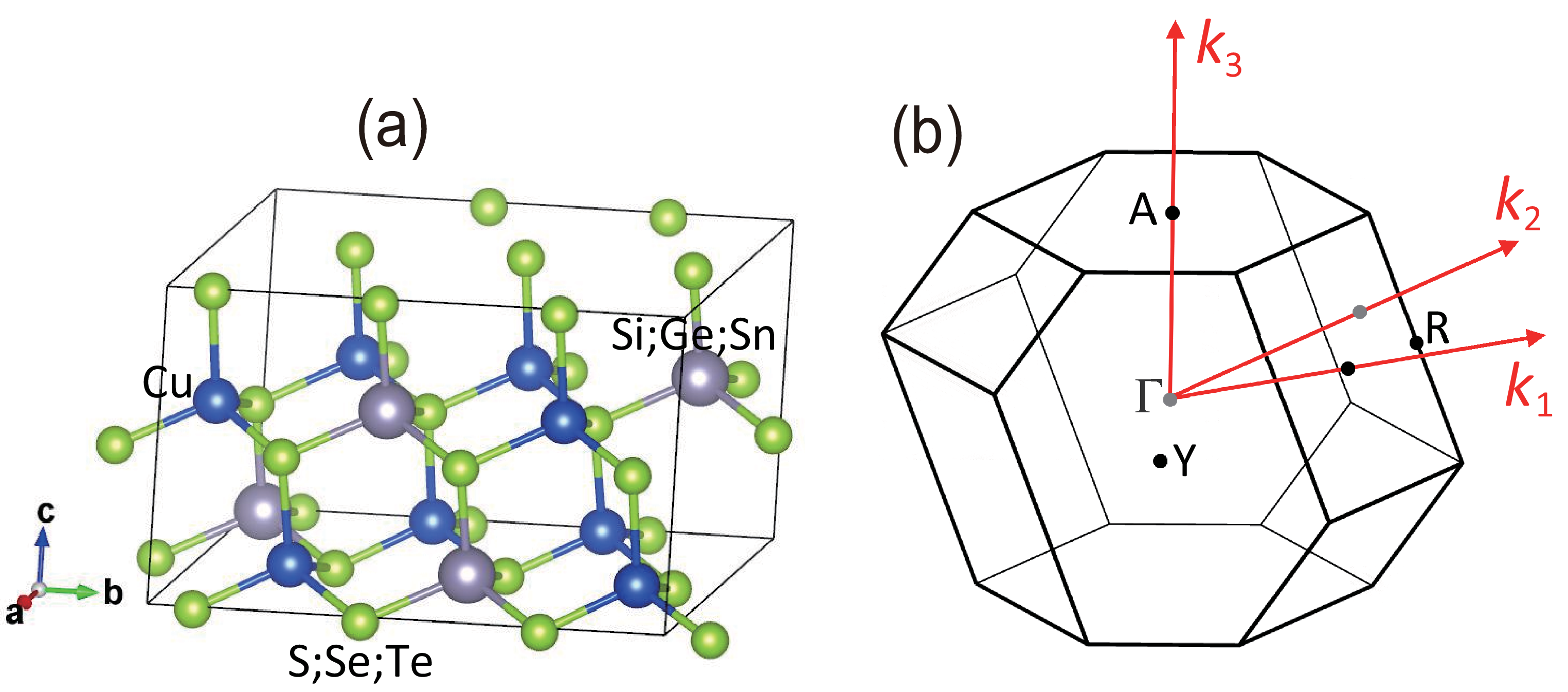}
\caption{ (a) Crystal structure of Cu$_2$SnSe$_3$ series materials with $Cc$ space group. (b) Corresponding BZ and high-symmetry points.
}
\label{Lattice}
\end{figure}

The Cu$_2$SnSe$_3$ series comprises the nine materials listed in Tab. \ref{tab1}. Their crystallographic unit cells are illustrated in Fig. \ref{Lattice}(a), while the corresponding BZ and high-symmetry points are depicted in Fig. \ref{Lattice}(b). These compounds adopt an enargite-like crystal structure and crystallize in the monoclinic system with space group $Cc$ (No. 9)~\cite{Delgado03,Amiri12,Vymazalova24}. It is important to note that the crystal structures of these materials lack inversion symmetry, and feature a glide mirror plane $M_{\mathbf{ac}}$ with a glide vector equals $\frac{1}{2}\mathbf{c}$. Taking Cu$_2$SnSe$_3$ as a representative example, the structure features two inequivalent Cu$^{2+}$ sites. At the first Cu$^{2+}$ site, the ion is coordinated by four Se$^{2-}$ anions to form a CuSe$_4$ tetrahedron, which shares vertices with five equivalent SnSe$_4$ tetrahedra and seven CuSe$_4$ tetrahedra. The Cu-Se bond lengths at this site range from 2.42 to 2.46 {\AA}. At the second Cu$^{2+}$ site, a similar CuSe$_4$ tetrahedron is formed, exhibiting an identical connectivity pattern - namely, vertex-sharing with five equivalent SnSe$_4$ and seven CuSe$_4$ tetrahedra - with Cu-Se bond distances spanning 2.42 to 2.47 {\AA}. The Sn$^{2+}$ ion occupies the center of a SnSe$_4$ tetrahedron defined by four Se$^{2-}$ ions, and this tetrahedron shares vertices with two equivalent SnSe$_4$ tetrahedra and ten CuSe$_4$ tetrahedra, resulting in Sn-Se bond lengths between 2.59 and 2.69 {\AA}. The structure further contains three inequivalent Se$^{2-}$ sites: at the first site, Se$^{2-}$ is coordinated to two Cu$^{2+}$ and two equivalent Sn$^{2+}$ ions, forming corner-sharing SeCu$_2$Sn$_2$ tetrahedra; at the second and third Se$^{2-}$ sites, each Se$^{2-}$ bonds to three Cu$^{2+}$ and one Sn$^{2+}$ ion, giving rise to corner-sharing SeCu$_3$Sn tetrahedra.

To accurately determine their electronic structures, we performed first-principles calculations based on density functional theory (DFT) using the Vienna ab-initio simulation package (VASP)~\cite{Kresse96}, employing the projector augmented wave (PAW) method and the generalized gradient approximation (GGA) with the Perdew-Burke-Ernzerhof (PBE) functional for the description of electron exchange-correlation effects. BZ integration was carried out using a 10$\times$10$\times$10 $\mathbf{k}$-point grid for the unit cell. SOC was included in all calculations unless otherwise stated. To validate the reliability of our computational results, we further performed band structure calculations using the WIEN2K package, which implements the full-potential linearized augmented plane wave (FP-LAPW) method~\cite{Blaha20}.
To systematically investigate the topological properties of these materials, we first performed DFT calculations in conjunction with the Wannier90 package to construct maximally localized Wannier functions~\cite{Pizzi20}. The tight-binding Hamiltonian was then built by fitting the Cu 3$d$ and S/Se/Te $p$ orbitals which dominate the electronic states near the Fermi level, ensuring accurate reproduction of the DFT band structure. This Hamiltonian was subsequently analyzed using the WannierTools package~\cite{Wu17} to systematically identify the $Z_2$ topological invariant, locate band node positions, compute the Berry curvature distribution, and calculate the corresponding topological surface states.

\heavyrulewidth=1bp

\begin{table}
\small
\renewcommand\arraystretch{1.3}
\caption{\label{tab1}
The parameters of Cu$_2$SnSe$_3$ series materials obtained from DFT calculations. The 2nd column indicates the presence or absence of a global gap in the absence of SOC; $\Delta_0$ and $\Delta_{soc}$ denote the gap values without and with SOC, respectively; $\lambda_{soc}$ represents the SOC strength of the S/Se/Te $p$ orbitals, as calculated using the TBSOC package~\cite{Gu23}; and the last column specifies the material's phase when SOC effects are included. All quantities are in units of eV.
}
\begin{tabular*}{8.9cm}{@{\extracolsep{\fill}}ccccccccc}
\toprule
 &   global gap?   &   $\Delta_0$   &   $\Delta_{soc}$      &   $\lambda_{soc}$  & phase \\
\hline
Cu$_2$SiS$_3$  & yes  &    1.09      & 1.08  &   0.10  & semiconductor  \\
Cu$_2$SiSe$_3$ & yes  &   0.36    &  0.33  &  0.23  & semiconductor  \\
Cu$_2$SiTe$_3$ & yes  &  0.24  & 0.10 &  0.60  & semiconductor  \\
Cu$_2$GeS$_3$  & yes  &  0.22  & 0.20 &   0.10 & semiconductor \\
Cu$_2$GeSe$_3$ & yes  & 0.045   & 0 &    0.22  & WSM \\
Cu$_2$GeTe$_3$ & no   &  -  & - &   0.55 & WSM \\
Cu$_2$SnS$_3$  & yes  & 0.018  & 0  &  0.10 & WSM \\
Cu$_2$SnSe$_3$ & yes  & 0.04   & 0 & 0.21 & WSM \\
Cu$_2$SnTe$_3$ & no   &  -  & - &  0.56   & WSM \\
\bottomrule
\end{tabular*}
\label{tab1}
\end{table}

\begin{figure*}[tbp]
\hspace{0cm} \includegraphics[totalheight=4.8in]{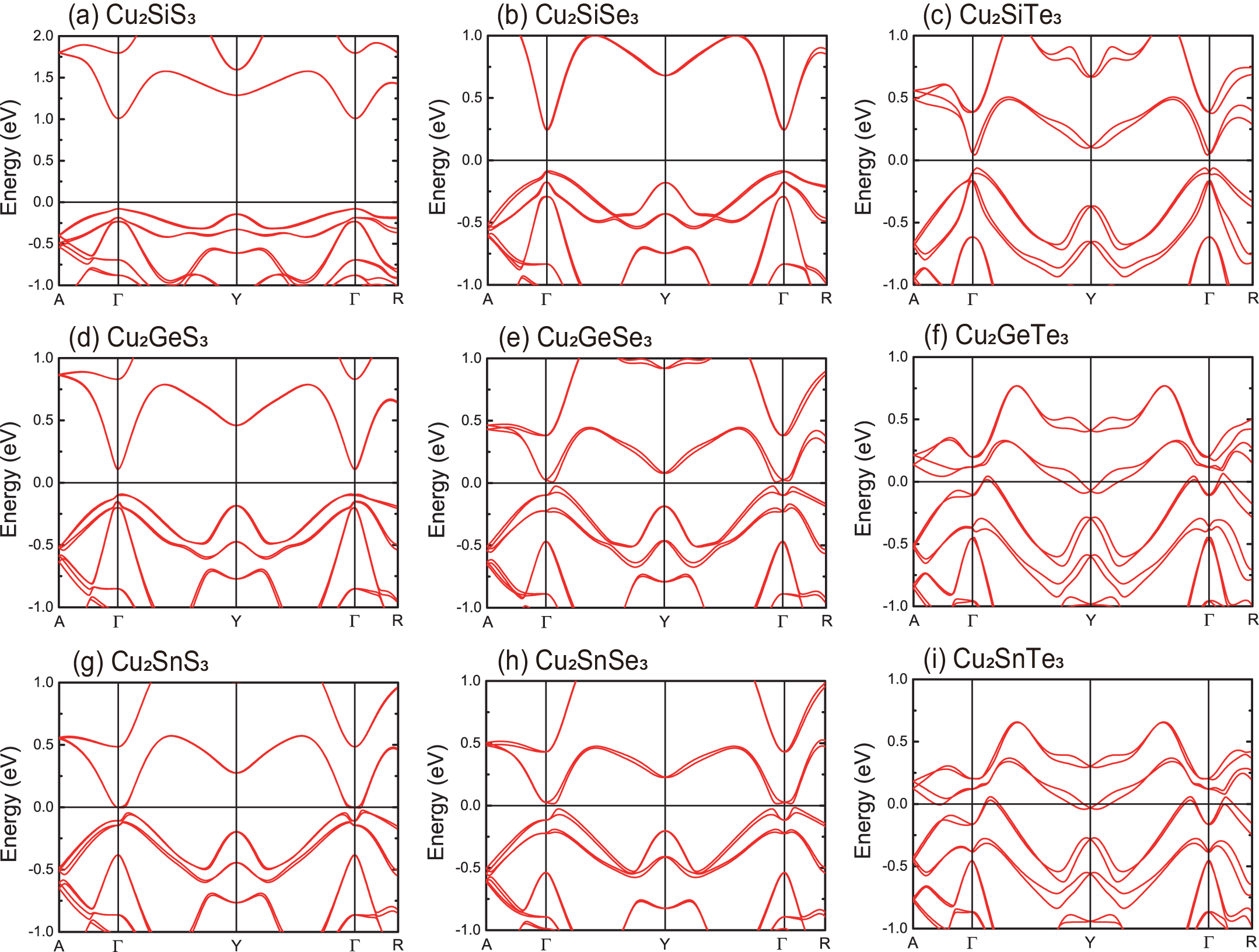}
\caption{ DFT band structures of nine materials along high-symmetry paths, with SOC effects fully included.
}
\label{allbands}
\end{figure*}

Fig. \ref{allbands} presents the band structures of these materials. The states near the Fermi level are predominantly derived from Cu 3$d$ and S/Se/Te $p$ orbitals, whereas the contributions from Si, Ge, and Sn in this energy region are negligible. Specifically, Cu$_2$SiS$_3$, Cu$_2$SiSe$_3$, Cu$_2$SiTe$_3$, and Cu$_2$GeS$_3$ exhibit relatively small global indirect band gaps.
For these insulating phases lacking SIS, their topological nature can be characterized by the $Z_2$ topological invariant~\cite{Kane05,Zhou22}. WannierTools calculations reveal that all these four materials have a $Z_2$ invariant of $(0; 000)$, confirming their classification as trivial semiconductors. Analysis of the band evolution across the nine compounds reveals that the band gap decreases significantly as the anion changes from S to Se to Te (left to right in Fig. \ref{allbands}), and similarly diminishes as the group-IV element progresses from Si to Ge to Sn (top to bottom in the figure). This trend is primarily attributed to the progressive elongation of the Cu-S, Cu-Se, and Cu-Te bonds and the concurrent expansion of the unit cell volume.

In Cu$_2$GeSe$_3$, Cu$_2$SnS$_3$, and Cu$_2$SnSe$_3$, the energy separation between the valence band maximum and conduction band minimum along high-symmetry $\mathbf{k}$-paths is extremely small, though whether a true gap closure occurs requires verification via higher-resolution calculations. In contrast, for Cu$_2$GeTe$_3$ and Cu$_2$SnTe$_3$, both valence and conduction bands cross the Fermi level, indicating the absence of a global band gap.

\begin{figure}[tbp]
\hspace{-0cm} \includegraphics[totalheight=2.75in]{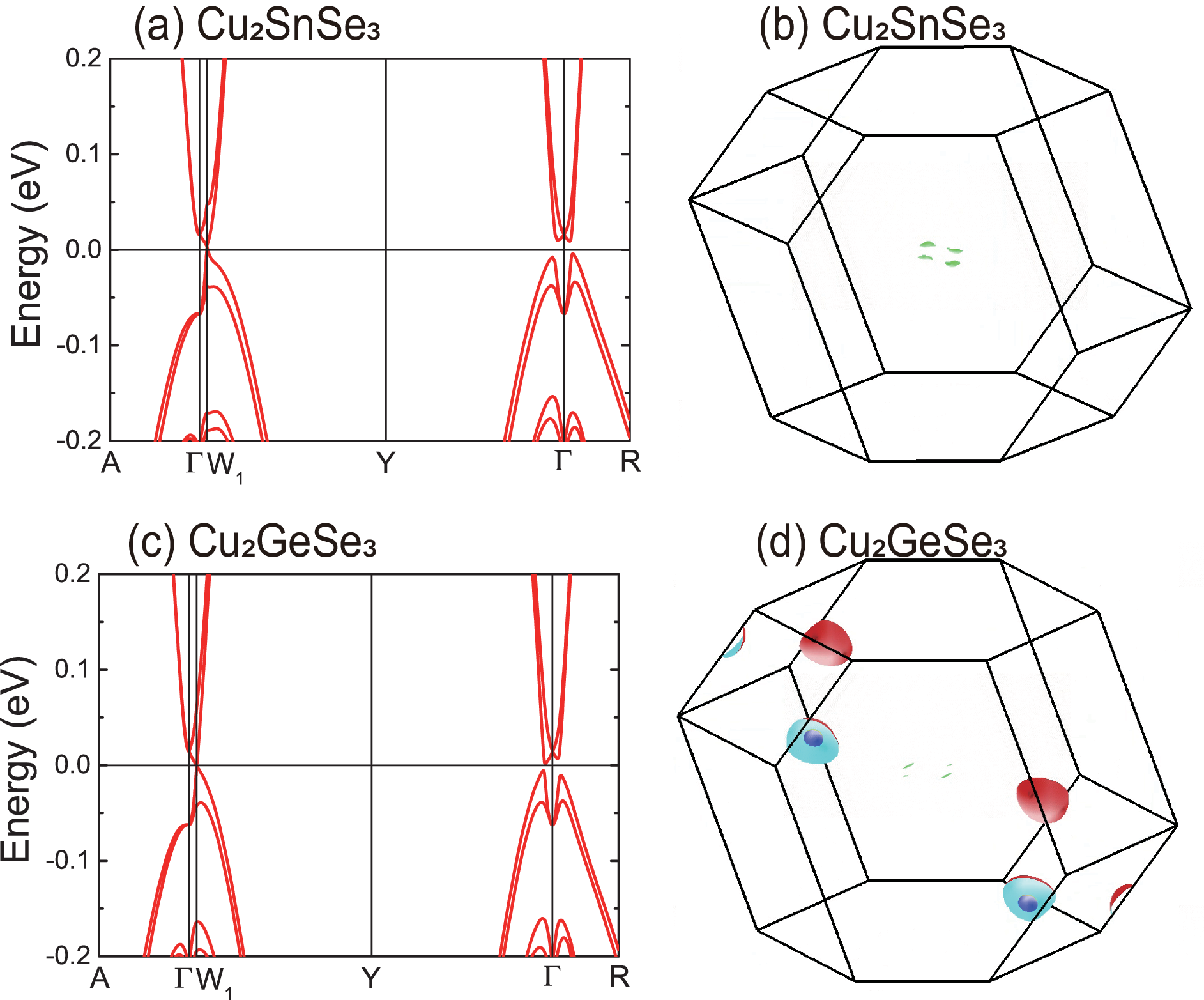}
\caption{ DFT band structures of Cu$_2$SnSe$_3$ and Cu$_2$GeSe$_3$ along one Weyl point W$_1$ (left column) and their corresponding Fermi surfaces (right column), with SOC included. Both materials exhibit four point-like Fermi surfaces near the BZ center (computed at 5 meV below the Fermi energy to enable clearer visualization), each arising from a Weyl cone.
}
\label{bands-alongWP}
\end{figure}

Notably, due to the lack of SIS in these crystal structures, SOC induces spin splitting of the bands, with the magnitude of splitting closely correlated to the SOC strength. Therefore, inclusion of SOC is essential for accurate determination of the band gap. As shown in Fig. \ref{allbands}, the closest approach between valence and conduction bands occurs near the $\Gamma$ point. To precisely resolve the gap at this location, we performed DFT calculations using a dense $\mathbf{k}$-mesh around the $\Gamma$ point, and the resulting gap values are summarized in Tab. \ref{tab1}. The third and fourth columns list the calculated gaps without and with SOC (all band gaps exhibit weak indirect character), respectively, demonstrating a substantial reduction upon SOC inclusion. To quantify the SOC strengths, we employed the TBSOC package~\cite{Gu23}, which confirms that the SOC effect on Cu 3$d$ is negligible, while the SOC parameters for $p$ orbitals in S, Se, and Te increase monotonically, with values of approximately 0.1 eV, 0.21-0.23 eV, and 0.55-0.60 eV, respectively, as listed in the fifth column of the table. The increasing SOC strength from S, Se to Te is consistent with the progressively enhanced spin splitting of the energy bands observed from left to right in Fig. \ref{allbands}. However, it should be emphasized that the spin splitting is significantly smaller than the intrinsic SOC strength of S/Se/Te, due to the hybrid character of the bands near the Fermi level, which arise from contributions of both the S/Se/Te $p$ and Cu 3$d$ orbitals.

These results confirm that Cu$_2$SiS$_3$, Cu$_2$SiSe$_3$, Cu$_2$SiTe$_3$, and Cu$_2$GeS$_3$ are intrinsic semiconductors. In contrast, Cu$_2$GeSe$_3$, Cu$_2$SnS$_3$, and Cu$_2$SnSe$_3$ exhibit vanishing band gaps and display linear band crossings near the $\Gamma$ point, forming Weyl points and thereby realizing a WSM phase (further discussed in the next section). The band dispersions around these Weyl points for Cu$_2$GeSe$_3$ and Cu$_2$SnSe$_3$ are shown in Figs. \ref{bands-alongWP}(a) and (c), respectively. The Weyl points lie extremely close to the Fermi level (within a few meV), giving rise to four isolated Fermi surface spots formed by Weyl cones, as illustrated in Figs. \ref{bands-alongWP}(b) and (d). In Cu$_2$SnSe$_3$, no other portions of the valence or conduction bands cross the Fermi level; hence, its Fermi surface consists solely of these four Weyl cone features. In Cu$_2$GeSe$_3$, however, additional crossings at the Fermi level occur away from the Weyl cones, leading to small electron and hole pockets on the Fermi surface. Although Cu$_2$GeTe$_3$ and Cu$_2$SnTe$_3$ also host Weyl points at four momentum-space locations, their extensive Fermi surfaces - resulting from broad valence and conduction band crossings at the Ferm level - render them less suitable for focused analysis in the present study.

\section{Weyl states in $\mathrm{Cu}_2\mathrm{SnSe}_3$ and $\mathrm{Cu}_2\mathrm{GeSe}_3$}

\begin{figure}[tbp]
\hspace{-0cm} \includegraphics[totalheight=2.4in]{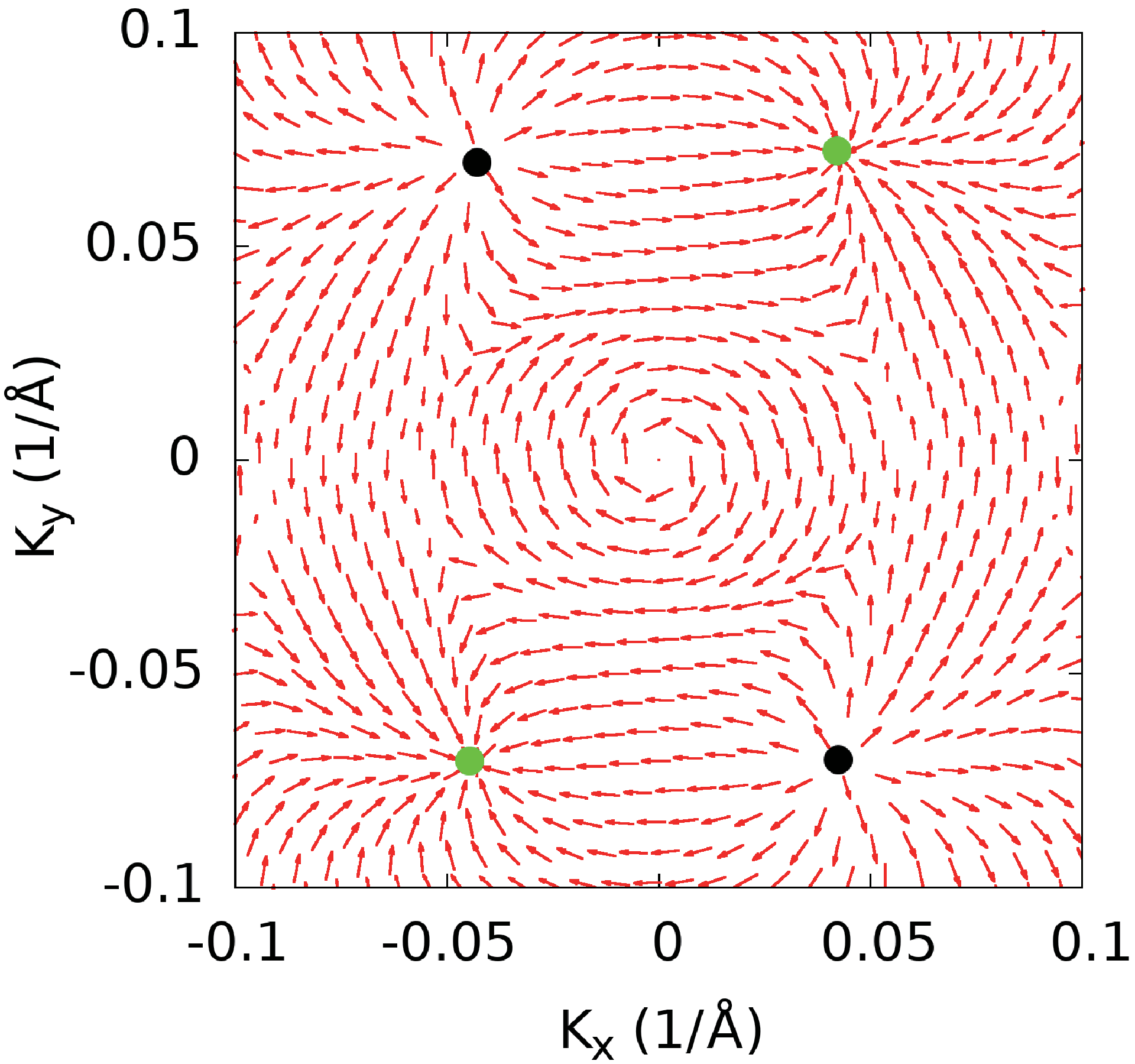}
\caption{ The Berry curvature of Cu$_2$SnSe$_3$ in the $\mathbf{k}$ plane contains four Weyl points, projected onto the $k_x$ - $k_y$ plane. The black and green dots denote Weyl points with chirality $+$1 and $-$1, respectively.
}
\label{Berry}
\end{figure}

As previously discussed, under the influence of SOC, the original indirect band gaps near the $\Gamma$ point in both Cu$_2$GeSe$_3$ and Cu$_2$SnSe$_3$ materials vanish, giving rise to four linear band crossings between the valence and conduction bands.
To determine whether these crossings correspond to Weyl points, we utilized the Wannier90 package to construct a tight-binding Hamiltonian that accurately reproduces the DFT band structure by fitting the Cu 3$d$ and Se 4$p$ orbitals that dominate the states near the Fermi level. This Hamiltonian was subsequently analyzed using the WannierTools package~\cite{Wu17} to systematically identify node positions, compute the Berry curvature distribution, and calculate topological surface states.
Our results reveal four nodal points in Cu$_2$SnSe$_3$ located at (kx, ky, kz) = (0.043, $\pm$0.074, -0.01) and (-0.043, $\pm$0.074, 0.01)${\AA}^{-1}$, in agreement with the DFT-derived positions. As shown in Fig. \ref{Berry}, the projection of the Berry curvature onto the $k_x$ - $k_y$ plane clearly reveals two source-like and two drain-like features, corresponding respectively to Weyl points of chirality $+$1 and $-$1.

\begin{figure*}[tbp]
\hspace{0cm} \includegraphics[totalheight=3.6in]{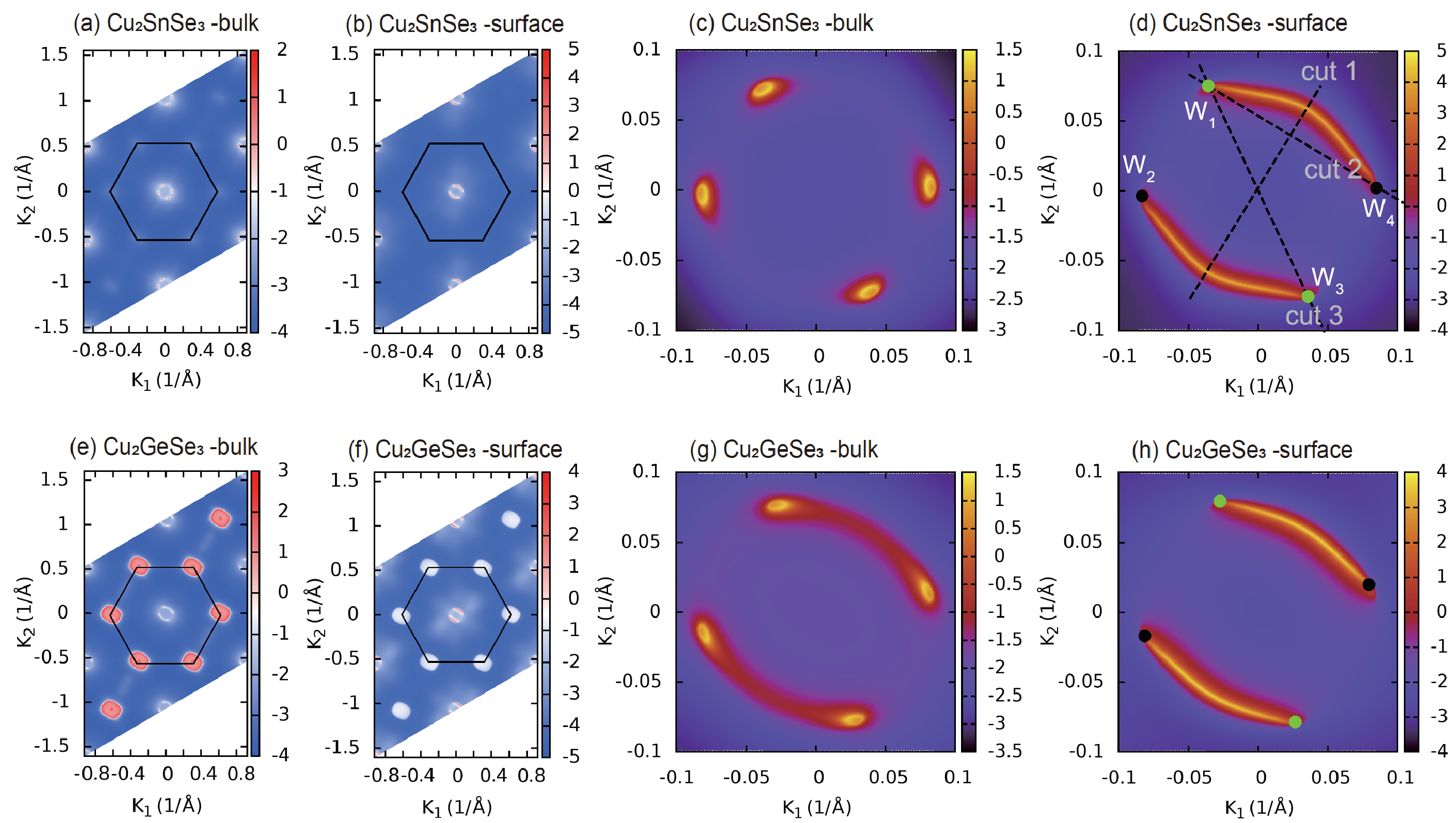}
\caption{ The first and second rows display the surface Fermi surfaces of Cu$_2$SnSe$_3$ and Cu$_2$GeSe$_3$ with opened (001) surface, while the first and second columns showing the contributions from bulk layers and surface layers, respectively. The two rightmost columns present corresponding magnified views.
}
\label{arc}
\end{figure*}

The four Weyl points in Cu$_2$SnSe$_3$ lie very close to the Fermi level, with energies deviating by only a few meV, resulting in nearly point-like Fermi surfaces. No additional Fermi surfaces are observed elsewhere in the BZ [see Fig. \ref{bands-alongWP}(b)], indicating that this system constitutes a highly pristine WSM phase, which is expected to host well-defined surface Fermi arcs. Fig. \ref{arc} presents the calculated surface Fermi surface distribution of the Cu$_2$SnSe$_3$ (001) surface - specifically, the Se-terminated surface - obtained using WannierTools, in which \ref{arc}(a) and \ref{arc}(b) display the contributions from bulk states and surface states, respectively. Fig. \ref{arc}(a) reveals point-like features near the center of the surface BZ (the hexagonal region), originating from the four Weyl cones [enlarged in Fig. \ref{arc}(c)]. In contrast, Fig. \ref{arc}(b) clearly shows two distinct Fermi arcs connecting these Weyl points [enlarged in Fig. \ref{arc}(d)]. A closer inspection of Fig. \ref{arc}(d) confirms that bright arc-like states link pairs of Weyl points with opposite chiralities - namely, $W_1$ and $W_4$, as well as $W_2$ and $W_3$. For Cu$_2$GeSe$_3$, the (001) surface Fermi surface exhibits similar characteristics (see second row of Fig. \ref{arc}): point-like bulk-derived features [Fig. \ref{arc}(g)] and two prominent Fermi arcs [Fig. \ref{arc}(h)]. However, additional electron and hole pockets appear at the corners of the BZ [see Fig. \ref{arc}(e)], which correspond to the projection of the bulk Fermi surface onto the (001) surface as shown in Fig. \ref{bands-alongWP}(d).

\begin{figure}[tbp]
\hspace{-0cm} \includegraphics[totalheight=1.9in]{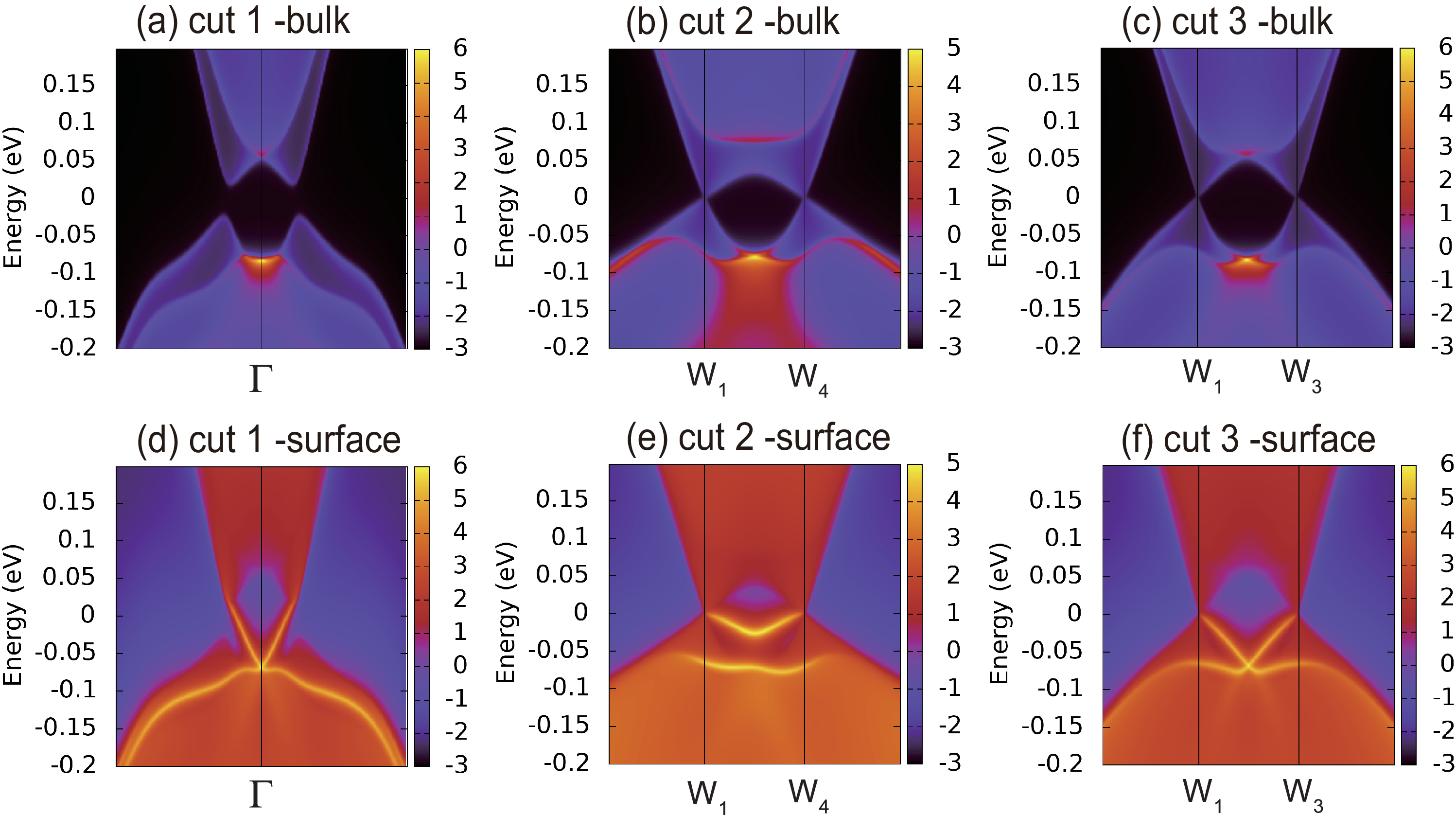}
\caption{ The band dispersion of Cu$_2$SnSe$_3$ with opened (001) surface. The first and second rows represent the contributions from bulk and surface layers, respectively. The three $\mathbf{k}$ cuts used in the plot correspond to the black dashed lines indicated in Fig. \ref{arc}(d).
}
\label{surfacestates}
\end{figure}

Fig. \ref{surfacestates} presents the surface-state dispersions of Cu$_2$SnSe$_3$ along three $\mathbf{k}$ paths within the surface BZ, as defined in Fig. \ref{arc}(d). Cut 1 does not pass through any Weyl points, cut 2 connects a pair of Weyl points with opposite chiralities, and cut 3 links two Weyl points of the same chirality. The bulk state dispersion along cut 1 [Fig. \ref{surfacestates}(a)] reveals a clear energy gap between the valence and conduction bands, while its surface state dispersion exhibits two bright in-gap states that traverse this bulk gap [Fig. \ref{surfacestates}(d)]. Along cut 2 [Fig. \ref{surfacestates}(b)], well-defined Weyl cones are observed at $W_1$ and $W_4$ - these arise from the projection of the three-dimensional bulk Weyl cones onto the (001) surface - and a prominent surface state connects these two points with opposite chiralities [Fig. \ref{surfacestates}(e)], consistent with the Fermi arc structure shown in Figs. \ref{arc}(d). For cut 3 [Fig. \ref{surfacestates}(c)], the bulk dispersion still displays sharp Weyl cones, yet no connecting surface state is observed between $W_1$ and $W_3$, which share the same chirality [Fig. \ref{surfacestates}(f)]. The dispersion features across all three cuts are in accordance with the Fermi arc maps [Figs. \ref{arc}(c) and \ref{arc}(d)]. Moreover, the surface state dispersions of Cu$_2$GeSe$_3$ exhibit nearly identical characteristics to those of Cu$_2$SnSe$_3$ and are therefore not further discussed here.

\section{spin-orbital coupling driven gap-closing}

\begin{figure*}[tbp]
\hspace{0cm} \includegraphics[totalheight=2.1in]{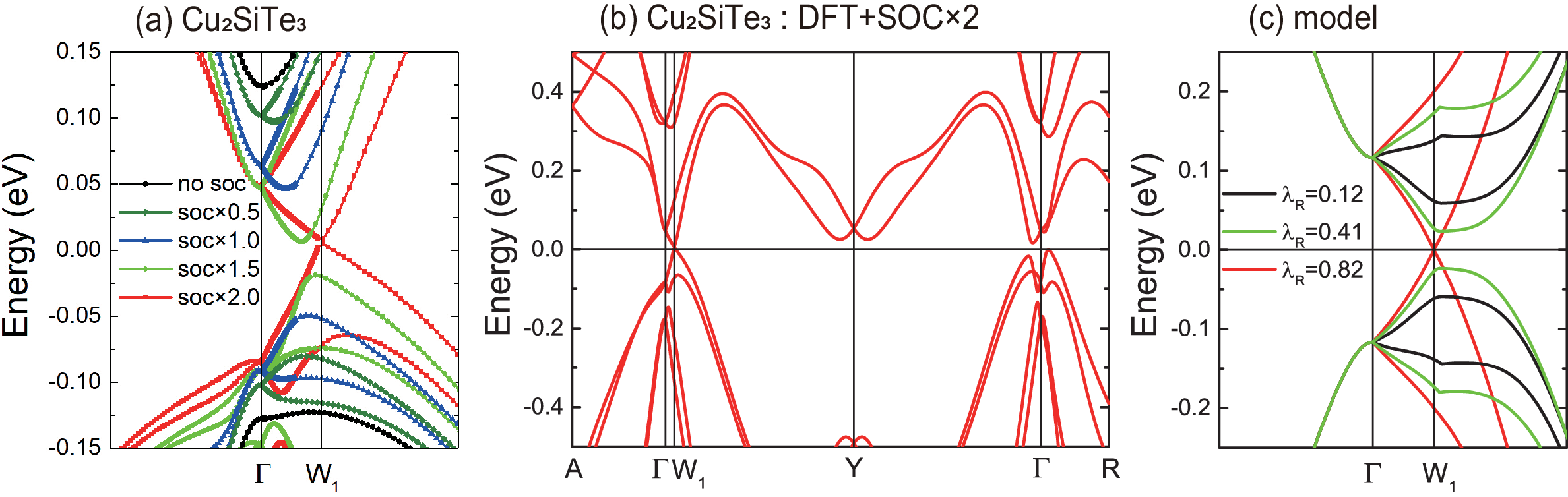}
\caption{(a) The evolution of the band structure in Cu$_2$SiTe$_3$ as a function of the SOC strength multiple. (b) The band structure passing through the Weyl point $W_1$ when the SOC strength is doubled; (c) The evolution of the bands with SOC strength $\lambda_R$ in the four-band model. The remaining parameters are set as $m=0.24$ eV$^{-1}$, $\Delta=0.06$ eV, $a_x=0.2$ eV, $a_y=2.0$ eV, $\xi=0.1$ eV, and $w=0.8$ eV in Eq. \ref{eq1}.
}
\label{Cu2SiTe3}
\end{figure*}

As previously discussed, the phase behavior of Cu$_2$SnSe$_3$ series materials under SOC - whether they remain in a semiconductor state or transition into a WSM phase - depends critically on the relative magnitude between the intrinsic band gap (i.e., the gap in the absence of SOC) and the strength of SOC. When the intrinsic band gap is significantly larger than the SOC strength, the system remains in the semiconductor phase, as exemplified by Cu$_2$SiS$_3$, Cu$_2$SiSe$_3$, and Cu$_2$GeS$_3$. In contrast, when the intrinsic band gap is small, even moderate SOC can drive the system into the WSM phase, as observed in Cu$_2$SnS$_3$, Cu$_2$SnSe$_3$, and Cu$_2$GeSe$_3$. A particularly illustrative case is Cu$_2$SiTe$_3$, which has an intrinsic band gap of 0.24 eV. Under a SOC strength of 0.6 eV, this gap is reduced to 0.1 eV, and further enhancement of SOC may lead to complete gap closure, thereby triggering a transition to the WSM state. Although the SOC strength in real materials cannot be externally tuned at will, DFT calculations allow for artificial scaling of the SOC parameter to probe the feasibility of such a TPT. Fig. \ref{Cu2SiTe3}(a) presents the evolution of the electronic bands near the indirect gap position in Cu$_2$SiTe$_3$ as the SOC strength is gradually increased from zero to twice its physical value. The results show a continuous reduction of the semiconductor band gap with increasing SOC, culminating in a linear band crossing at twice the nominal SOC strength. This crossing forms a characteristic Weyl cone structure, signaling the emergence of the WSM phase. These findings confirm that tuning the SOC strength alone is sufficient to induce a TPT from a semiconductor to a WSM.

This transition can be effectively described using a low-energy model. By selecting one valence band and one conduction band near the Fermi level, the inclusion of SOC leads to their splitting into four distinct bands, which can be captured by a four-band Hamiltonian near $\Gamma$ point~\cite{Chiu14,Deaconu25}
\begin{align}\mathcal{H}(\mathbf{k})=&(\frac{\mathbf{k}^2}{m}-\Delta)\tau_3\otimes \sigma_0+a_xk_x\tau_1\otimes\sigma_2+a_zk_z\tau_2\otimes\sigma_0\nonumber\\+&\xi\tau_1\otimes\sigma_0
+\lambda_R\tau_1\otimes(k_x\sigma_2-k_y\sigma_1)\nonumber\\+&w(k^3_x-3k_xk^2_y)\tau_2\otimes\sigma_0.
\label{eq1}
\end{align}
The first term accounts for the parabolic dispersion near the $\Gamma$ point, while the second and third terms represent linear momentum dependence. The fourth and fifth terms correspond to the mass term and Rashba-type SOC, respectively, which break SIS and are essential for stabilizing the WSM phase. The final term describes the warping effect on the Fermi surface, which arises due to the lattice glide mirror symmetry. Using parameters appropriate for Cu$_2$SiTe$_3$, we compute the band structure evolution as a function of SOC strength [see Fig. \ref{Cu2SiTe3}(c)]. The calculations reveal that the band gap begins to close at a scaled SOC strength of $\lambda_R=0.648$ eV, accompanied by the emergence of four Weyl points - clear evidence of an SOC-driven semiconductor-to-WSM TPT.

\section{topological phase transitions under doping}

\begin{figure}[tbp]
\hspace{0cm} \includegraphics[totalheight=3.05in]{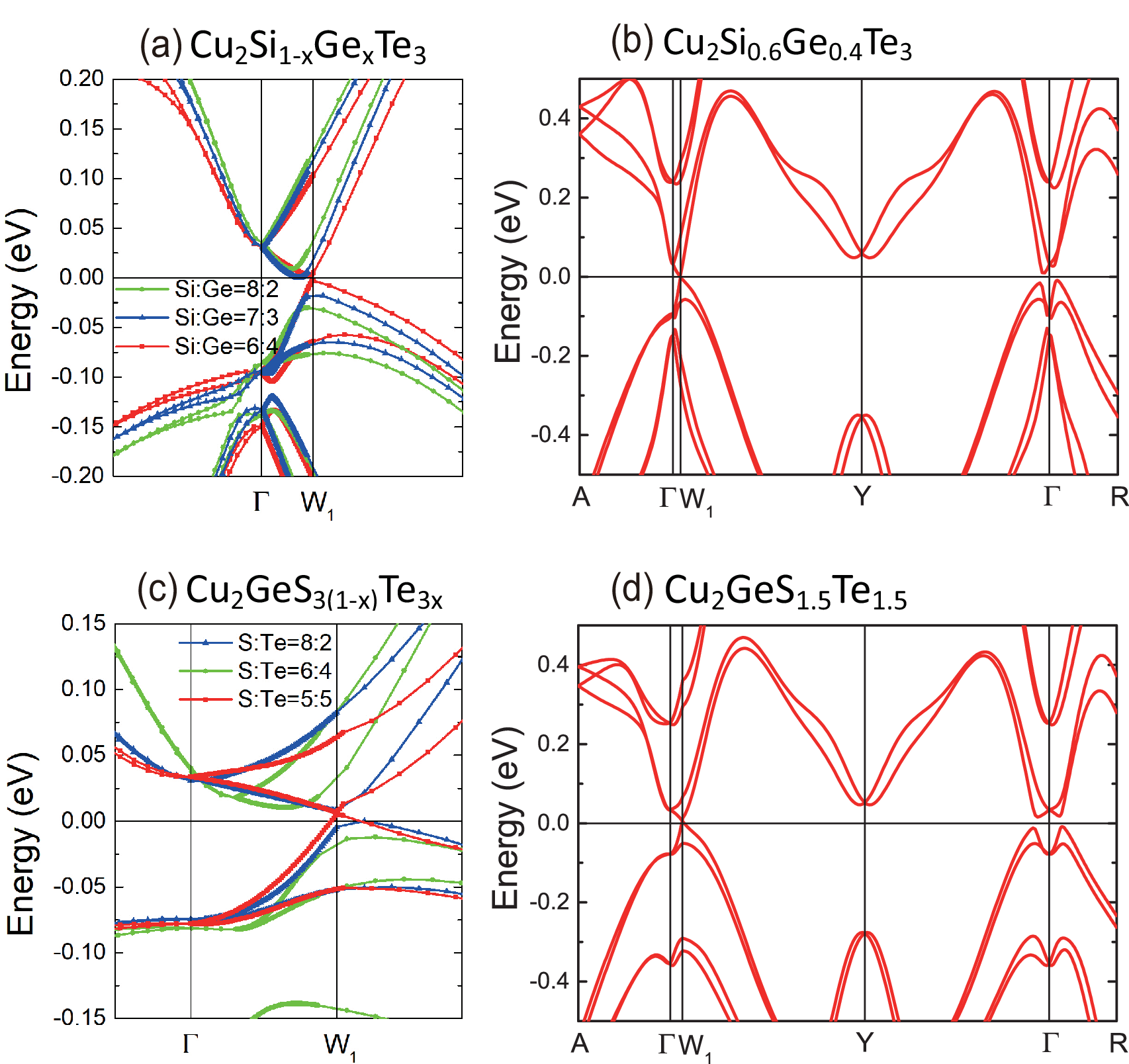}
\caption{ The left panels display the evolution of the band structures for the two doped systems as functions of doping concentration, while the right panels illustrate the band dispersions crossing the Weyl point $W_1$ in the WSM phases.
}
\label{doping}
\end{figure}

According to the foregoing analysis, achieving the TPT from an insulating state to a WSM fundamentally requires the closure of the bulk energy gap to enable the emergence of Weyl points. This can be realized either by reducing the bulk gap or by enhancing spin splitting induced by SOC. As shown in the first column of the band structures for the nine materials in Fig. \ref{allbands}, the bulk gap decreases significantly when transitioning from Si to Ge and then to Sn, primarily due to the progressive enlargement of the unit cell volume. Similarly, a pronounced reduction in the bulk gap is also observed along the first row - from S to Se to Te - attributable to the increasing lengths of the Cu-S, Cu-Se, and Cu-Te bonds as well as the expanding unit cell size, both of which collectively reduce the energy splitting between bonding and antibonding states.

Therefore, to realize the TPT from a semiconductor to a WSM, one may select material systems with intrinsically small bulk gaps and further drive gap closure through the aforementioned tuning strategies. Band structures in Fig. \ref{allbands} reveal that Cu$_2$SiTe$_3$ and Cu$_2$GeS$_3$ exhibit relatively small bulk gaps, making them promising candidate parent compounds for investigating such transitions.

In the Cu$_2$SiTe$_3$ system, partial substitution of Si with Ge effectively reduces the energy gap. As illustrated in Fig. \ref{allbands}(f), Cu$_2$GeTe$_3$ exhibits no global band gap and displays signatures of Weyl points (see Tab. \ref{tab1}), indicating that even partial replacement of Si by Ge is sufficient to induce gap closure. Experimentally, this can be realized via Ge doping at the Si sites in Cu$_2$SiTe$_3$. For Cu$_2$GeS$_3$, although its initial band gap is somewhat larger, the fact that Cu$_2$GeTe$_3$ already possesses a closed gap suggests that Te doping at the S sites can simultaneously reduce the band gap and enhance SOC strength, thereby synergistically driving the system toward gap closure.

The electronic structures of the two doped systems Cu$_2$Si$_{1-x}$Ge$_x$Te$_3$ and Cu$_2$GeS$_{3(1-x)}$Te$_{3x}$ were simulated using the virtual crystal approximation (VCA) within the framework of DFT. Lattice constants and atomic positions were determined through linear interpolation of structural parameters between the parent compound ($x$=$0$) and the end member ($x$=$1$), based on doping concentration $x$~\cite{Zhou22}. The resulting band structures are presented in Fig. \ref{doping}. The left panels depict the evolution of the bands along the gap positions with increasing doping level: the system exhibits an indirect band gap, and with rising $x$, the valence band maximum and conduction band minimum progressively converge in $\mathbf{k}$-space before crossing to form Weyl points. Calculations indicate that the band gaps close at $x$=$0.4$ for Cu$_2$Si$_{1-x}$Ge$_x$Te$_3$ and $x$=$0.5$ for Cu$_2$GeS$_{3(1-x)}$Te$_{3x}$, coinciding with the emergence of Weyl points. The linear dispersion relations around these points, displayed in the right panels of Fig. \ref{doping}, further confirm the presence of Weyl fermions.

In conclusion, chemical doping provides an effective route to drive the TPT from semiconductor to WSM. This mechanism does not rely on structural distortions, external strain, or magnetic ordering, but instead operates solely through compositional tuning, thus offering high experimental feasibility and excellent controllability.

\section{conclusion}

In summary, based on systematic first-principles calculations, we demonstrate that the Cu$_2$SnSe$_3$ family of materials with non-centrosymmetric $Cc$ space group symmetry serves as an ideal platform for realizing a topological phase transition from a trivial semiconductor to a WSM. In this series of compounds, the electronic phase - whether semiconductor or WSM - is governed by the relative evolution between the intrinsic band gap and the strength of SOC, with the closure of the bulk band gap signaling the onset of the TPT. Specifically, Cu$_2$SnSe$_3$ and Cu$_2$GeSe$_3$ exhibit WSM characteristics, featuring nearly point-like Fermi surfaces predominantly contributed by four Weyl nodes, which give rise to well-defined Fermi arcs on the surface Brillouin zone, thus representing prototypical WSMs. The transition from the semiconductor to the WSM phase can be driven by either reducing the intrinsic band gap or enhancing SOC through elemental substitution: the former is exemplified by Si-to-Ge doping in Cu$_2$SiTe$_3$, while the latter is achieved by substituting S gradually with Te in Cu$_2$GeS$_3$. These chemically induced TPT preserve the crystal symmetry and do not require external strain or high-pressure tuning, making them highly accessible in experimental settings.

It should be emphasized that our calculations place Cu$_2$SnSe$_3$ and Cu$_2$GeSe$_3$ in close proximity to the semiconductor - WSM phase boundary. Therefore, their precise classification - as semiconductors or semimetals - requires more accurate lattice parameters, higher-resolution theoretical simulations, and ultimately direct experimental verification via measurements of the Fermi surface and surface states. Nevertheless, it is evident that a slight enhancement of SOC strength is sufficient to drive these compounds into the WSM phase, a scenario readily achievable through minimal Se-site Te doping. In addition, we note that another compound with a non-centrosymmetric structure, Cu$_2$SnS$_3$ (belonging to the $Imm2$ space group), has been identified as a WSM; in this system, Se substitution for S can drive a TPT from a WSM to a TI~\cite{Zhou22}. However, due to its relatively large bulk Fermi surface, the resulting surface Fermi surface structure becomes significantly more complex than that of the $Cc$-space-group compounds investigated in this study. Consequently, the Cu$_2$SnSe$_3$ series compounds proposed herein represent a particularly promising candidate system for ideal WSM, and the strategy of deriving WSM from semiconductor parent phases offers a valuable route toward realizing Weyl systems with larger $\mathbf{k}$-point separation between Weyl points.

\acknowledgments
H. Li acknowledgements the supports from National Natural Science Foundation of China (No. 12364023), and Guangxi Natural Science Foundation (No. 2024GXNSFAA010273).

\end{document}